\documentstyle[12pt]{article}
 
\newcommand{\doublespace}{\renewcommand{\baselinestretch}{1.75}
\Large\normalsize}
 
\begin{document}
\begin{titlepage}

\title{The Teleparallel Lagrangian and Hamilton-Jacobi Formalism}

\author{B. M. Pimentel$^{a}$,
P. J. Pompeia$^{a,b}$,\\
J. F. da Rocha-Neto$^{*,a}$ and R. G.
Teixeira$^{c}$\\ 
$^{a}$Instituto de F\'{\i}sica Te\'{o}rica\\
 - Universidade Estadual Paulista,\\
Rua Pamplona 145, 01405-900, S\~{a}o Paulo, SP, Brazil\\ 
$^{b}$Centro T\'{e}cnico Aeroespacial,\\ 
Intituto de Fomento e Coordena\c{c}\~{a}o Industrial,\\
Divis\~{a}o de Confiabilidade Metrol\'{o}gica, Pra\c{c}a Marechal\\ 
Eduardo Gomes 50, 12228-901,
S\~{a}o Jos\'{e} dos Campos, SP, Brazil \\
$^{c}$Faculdade de Tecnologia e Ci\^{e}ncias\\ 
Exatas - Universidade S\~{a}o Judas Tadeu,\\
Rua Taquari 546, 03166-000, S\~{a}o Paulo, SP, Brazil.\\}
\date{}
\maketitle

\begin{abstract}
We analyze the Teleparallel Equivalent of General Relativity (TEGR) from the
point of view of Hamilton-Jacobi approach for singular systems.
\end{abstract}

\thispagestyle{empty} 
\vfill
\par
\noindent (*) e-mail: rocha@ift.unesp.br\par
\end{titlepage}
\newpage
\doublespace
\section{Introduction}

The analysis of singular systems is an interesting problem in Physics, as such
systems appear in many relevant physical problems. Such analysis is usually
carried out using the generalized Hamiltonian formulation, developed by Dirac
\cite{Di, Sundermeyer, Tei, Git}, where the canonical Hamiltonian is not uniquely
determined due to the singularity of the Hessian matrix; what causes the
appearance of relations between canonical variables. These constraints,
multiplied by Lagrange multipliers, are added to the canonical Hamiltonian and
consistency conditions are implemented to eliminate some degrees of freedom of
the system.

Despite the outstanding success of Dirac's formalism, new approaches to the
analysis of singular systems are always welcome because they may reveal new
mathematical and physical information about the system in study. Among others,
an alternative method to analyze singular systems is the Hamilton-Jacobi
formalism \cite{G1, G2}, which has been used in many examples \cite{G3, G4,
G5} and generalized to higher order singular systems \cite{B1,B2} and systems
with Berezinian variables \cite{B3}. This formalism uses Carath\`{e}odory's
equivalent Lagrangians method \cite{Ca} to write a set of Hamilton-Jacobi
partial differential equations from which one can obtain the equations of
motion as a set of total differential equations in many variables.

One example of physical system described by a singular Lagrangian, and that
has already been studied through Dirac's method \cite{M94, MR99, Iuguslavia},
is the Teleparallel Equivalent of General Relativity (TEGR) which is an
alternative formulation of General Relativity \cite{H} developed in
Weitzenb\"{o}ck space-time \cite{W}. In oposition to General Relativity, in
TEGR the curvature tensor vanishes but the torsion tensor does not so that, in
this geometrical framework, the gravitational effects are caused by the
torsion tensor and not by curvature.

The TEGR has been successfully analyzed through Dirac's Hamiltonian formalism
\cite{M94, MR99} and generated successfull applications \cite{ME}. Moreover,
many of the characteristics regarding the interaction of spin 0, 1 and spinor
fields in TEGR have been studied recently \cite{Andrade 2, Andrade 3, Andrade
4, DKP 1, DKP 2} as well as its gauge symmetries \cite{Iuguslavia 2}. Others
aspects of TEGR, as the energy-momentum tensor and geodesics and
``force''\ equation are addressed in reference \cite{Andrade 1} and in the
references cited therein.

Our intention in this work is to add a different point of view to the analysis
of  TEGR by studying it through the above mentioned Hamilton-Jacobi formalism
for singular sistems. First, we will introduce the Lagrangian density of TEGR
in a form which is appropriate to our approach. Then we address the basic
aspects of Hamilton-Jacobi formalism for singular systems and apply such
formalism to TEGR. Finally, we present our final comments.

\section{The Lagrangian of TEGR}

In this section we summarize the Lagrangian formulation of TEGR in terms
of the tetrad field, as presented in reference \cite{MR99} where a global
$SO(3,1)$ symmetry is taken from outset. This choice is done because, when
starting from a local $SO(3,1)$ symmetry, it may not be possible, with
certainty reference systems choices, to obtain a set of first class
constraints \cite{M94}.

So, we take the Lagrangian density of TEGR in empty space-time given, in
terms of the tetrad field $e_{a\mu}$, by
\begin{equation}
L(e)=-k\,e\,\Sigma^{abc}T_{abc}, \label{2.1}
\end{equation}
where Latin letters are $SO(3,1)$ indexes (taking values $(0),...(3)$), Greek
letters are space-time ones (taking values $0,...,3$), $e=det(e^{a}\,_{\mu})$,
$k=\frac{1}{{16\pi G}}$ and $G$ is the gravitational constant. Besides that,
the torsion tensor $T_{abc}=e_{b}\,^{\mu}e_{c}\,^{\nu}T_{a\mu\nu}$ is defined
in terms of the tetrad field as
\[
T_{a\mu\nu}\;=\;\partial_{\mu}e_{a\nu}-\partial_{\nu}e_{a\mu},
\]
and its trace is defined as
\[
T_{b}=T^{a}\,_{ab};
\]
while the tensor $\Sigma^{abc}$ is defined as
\[
\Sigma^{abc}\;=\;{\frac{1}{4}}(T^{abc}+T^{bac}-T^{cab})+{\frac{1}{2}}
(\eta^{ac}T^{b}-\eta^{ab}T^{c}),
\]
such that
\[
\Sigma^{abc}T_{abc}\;=\;{\frac{1}{4}}T^{abc}T_{abc}+{\frac{1}{2}}
T^{abc}T_{bac}-T^{a}T_{a}.
\]

The fields equations can be obtained from the variation of $L$ with respect to
$e^{a\mu}$ and are equivalent to Einstein's equations in tetrad form \cite{M94}

\begin{equation}
\frac{{\delta L}}{{\delta e^{a\mu}}}\;\equiv\;\frac{1}{2}\,e\,\biggl\{R_{a\mu
}(e)-\frac{1}{2}e_{a\mu}R(e)\biggr\}. \label{2.2}
\end{equation}

Let us now consider the tetrad field $e^{a\mu}$ in terms of the $3+1$
decomposition. In this case the space-time manifold is assumed to be
topologically equivalent to $M\times R$, where $M$ is a noncompact
three-dimensional manifold. We consider that $^{4}e_{a\mu}$ is a tetrad field
for $M\times R$. In terms of the lapso $N$ and shift $N^{i}$
($i,j,k...=1,...,3$) functions we have
\[
^{4}e^{a}\;_{i}\;=\;e^{a}\,_{i},
\]
\[
^{4}e^{ai}\;=\;e^{ai}+(N^{i}/N)\eta^{a},
\]
\[
^{4}e^{a}\,_{0}\;=\;e^{a}\,_{i}N^{i}+\eta^{a}N,
\]
\[
e^{ai}\;=\;e^{a}\,_{k}{\bar{g}}^{ki},
\]
\[
\eta^{a}\;=\;-N\;^{4}e^{a0},
\]
where $e_{ai}$ and $e^{ai}$ are now restricted to $M$. Moreover, $\eta^{a}$ is
an unit timelike vector such that $\eta_{a}e^{a}\,_{i}=0$ and $\bar{g}^{ik}$
is the inverse of $g_{ik}=e_{ai}e^{a}\,_{k}$. The determinant of the tetrad
field is now given by $^{4}e\;=\;Ne$, where $e\;=\;det(e^{(l)}\,_{k})$. In terms
of the 3+1 decomposition, the TEGR Lagrangian density $L$ can be written as

$$L\;  =\;{\frac{ke}{2N}}\left(  \bar{g}^{ik}l_{i}^{(l)}l_{(l)k}
+e^{(0)i}e^{(0)k}l^{(l)}\,_{i}l_{(l)k}+e^{(l)i}e^{(n)k}l_{(n)i}l_{(l)k}
-2e^{(l)i}e^{(n)k}l_{(l)i}l_{(n)k}\right)$$
$$ -4ke\bar{\Sigma}^{a(0)i}l_{ai}-kNe\bar{\Sigma}^{abc}\bar{T}_{abc}$$

where we made the use of the following definitions
\[
l_{ai}\;=\;\dot{e}_{ai}-\eta_{a}\partial_{i}N-e_{aj}\partial_{i}N^{j}
-N^{j}\partial_{j}e_{ai},
\]
\[
\bar{\Sigma}^{abc}\;=\;{\frac{1}{4}}\left(  \bar{T}^{abc}+\bar{T}^{bac}
-\bar{T}^{cab}\right)  +{\frac{1}{2}}\left(  \eta^{ac}\bar{T}^{b}-\eta
^{ab}\bar{T}^{c}\right)  ,
\]
\[
\bar{T}^{abc}\;=\;e^{bi}e^{cj}T^{a}\,_{ij}.
\]
The momenta $\Pi^{ak}$ conjugated to $e_{ak}$ are obtained by
\[
\Pi^{ak}\;=\;{\frac{\delta L}{\delta\dot{e}_{ak}}},
\]
where
\[
\Pi^{(0)k}\;=\;4ke\bar{\Sigma}^{(0)(0)k},
\]
end
\[
\Pi^{(r)s}\;=\;-{\frac{ke}{N}}\left(  l^{(r)s}+e^{(0)i}e^{(0)s}l^{(r)}
\,_{i}+e^{(r)i}e^{(n)s}l_{(n)i}-2le^{(r)s}\right)  +4ke\bar{\Sigma}^{(r)(0)s}.
\]
Note that here $l^{(l)i}=\bar{g}^{ik}l^{(l)}\,_{k}$ and $l=e^{(l)i}l_{(l)i}$.

In order to simplify the calculations we can make a choice of reference frame,
analogous to the one made in ADM formulation. This choice is 
usually refered in literature as the Schwinger's time gauge \cite{Sch},
$e_{(0)i}\;=\;e_{(0)}\,^{i}=0$, so that the momenta defined above are
such that $\Pi^{ks}=\Pi^{sk}$ and $\Pi^{(0)k}=\bar{T}^{(l)}\,_{(l)}\,^{k}$. In
this case it is possible
to write the Lagrangian density $L$ as \cite{M94}
\begin{equation}
L\;=\;\Pi^{(l)i}\dot{e}_{(l)i}-H_{c},\label{2.3}
\end{equation}
where
\begin{equation}
H_{c}\;=\;NC+N^{i}C_{i},\label{2.4}
\end{equation}
\begin{equation}
C\;=\;{\frac{1}{4e}}\left(  \Pi^{ij}\Pi_{ij}-{\frac{1}{2}}\Pi^{2}\right)
+e\Sigma^{ijk}T_{ijk}-2\partial_{i}(eT^{i}),\label{2.5}
\end{equation}
and
\begin{equation}
C_{i}\;=\;e_{(l)i}\partial_{k}\Pi^{(l)k}+\Pi^{(l)k}T_{(l)ki}.\label{2.6}
\end{equation}

In Eq. (\ref{2.3}) there is no time derivatives of the functions $N$ and
$N^{i}$, so the Lagrangian density $L$ is singular. With these
results, we now can investigate the integrability conditions of
the Lagrangian density of the TERG, given in Eq.
(\ref{2.3}), in the Hamilton-Jacobi formalism, what it will be done in \ the
next section.

\section{The Hamilton-Jacobi Formalism}

The Hamilton-Jacobi formalism, recently developed to analyze singular systems
\cite{G1,G2,B1,B3}, uses the equivalent Lagrangians method \cite{Ca} to obtain
a set of Hamilton-Jacobi partial differential equations \cite{G1,G2}. We
suggest the references just mentioned for details and present here only the
main aspects of this formalism. For this, let us consider a singular
Lagrangian function $L=L(q_{i},\dot{q}_{i},t),$ where $i=1,...,N$. The Hessian
matrix is then given by
\begin{equation}
H_{ij}={\frac{\partial^{2}L}{\partial\dot{q}^{i}\partial\dot{q}^{j}}
}\;\;i,j=1,...,N. \label{3.1}
\end{equation}

Being the rank of the Hessian matrix $P=N-R<N$, we can define, without loss of
generalization, the order of the variables $q_{i}$ in a such way that the
$P\times P$ matrix in the right bottom corner of the Hessian matrix be
nonsingular. So there will be $R$ relations among canonical variables given
by
\begin{equation}
p_{\alpha}=-H_{\alpha}\left(  q^{i};\ p_{a}\right)  ;\ \alpha
=1,...,R;\label{a19}
\end{equation}
which correspond to Dirac's primary constraints $\Phi_{\alpha}\equiv
p_{\alpha}+H_{\alpha}\left(  q^{i};\ p_{a}\right)  \approx0$. From these we
get the Hamilton-Jacobi partial differential equations, given by
\begin{equation}
H_{\alpha}^{\prime}=H_{\alpha}(t,q_{i},p_{a})+p_{\alpha}=0,\label{3.2}
\end{equation}
\begin{equation}
H_{0}^{\prime}=H_{c}(t,q_{i},p_{a})+p_{0}=0,\label{3.3}
\end{equation}
where
\begin{equation}
p_{j}={\frac{\partial S(t,q_{i})}{\partial q^{j}}},\label{3.4}
\end{equation}
\begin{equation}
p_{0}={\frac{\partial S(t,q_{i})}{\partial t}},\label{3.5}
\end{equation}
and $i,j=1,...,N$, $\alpha=1,...,R$; $a=R+1,...,N$; $H_{c}$ is the canonical
Hamiltonian and $S$ is the Hamilton principal function.

It can be shown that the equations of motion are total differential equations
for the characteristics curves of the differential partial equations
(\ref{3.2}) and (\ref{3.3}), being given by \cite{G1}
\begin{equation}
dq_{i}=\frac{\partial H_{0}^{\prime}}{\partial p^{i}}dt+\frac{\partial
H_{\alpha}^{\prime}}{\partial p^{i}}dq_{\alpha},\label{3.5b}
\end{equation}
\begin{equation}
dp_{i}=-\frac{\partial H_{0}^{\prime}}{\partial q^{i}}dt-\frac{\partial
H_{\alpha}^{\prime}}{\partial q^{i}}dq_{\alpha},\label{3.5c}
\end{equation}
where for $i=1,...,R$ equation (\ref{3.5b}) above becomes a trivial identity.
Using standard techniques of partial differential equations, it can be shown
\cite{B3} that the equations above are integrable if and only if the functions
$H_{\alpha}^{\prime}$ satisfy
\begin{equation}
dH_{\beta}^{\prime}=\left\{  H_{\beta}^{\prime},H_{\alpha}^{\prime}\right\}
dt^{\alpha},\label{i14}
\end{equation}
where $\alpha,\beta=0,1,...,R$; $t_{0}=t$ (so that $t_{\nu}=(t,q_{\alpha})$)
and the symbol $\left\{  ...,...\right\}  $ denotes the Poisson bracket
defined on the phase space of $2N+2$ dimension that includes $t_{0}=t$ and its
canonical momentum $p_{0}$.

\subsection{The Teleparallel Lagrangian case}

Let us now consider the Lagrangian density given in Eq. (\ref{2.3}) in the
context of the Hamilton-Jacobi formalism. For this purpose we note that this
Lagrangian density does not depend on time derivatives of the shift and lapso
functions, therefore we define the set of Hamilton-Jacobi partial differential
equations as
\begin{equation}
H_{0c}^{\prime}=\int d^{3}x(H_{c}(x)+p_{0}(x))=0,\label{3.7}
\end{equation}
\begin{equation}
H_{0}^{\prime}=\int d^{3}x\Pi_{0}(x)=0,\label{3.8}
\end{equation}
\begin{equation}
H_{i}^{\prime}=\int d^{3}x\Pi_{i}(x)=0,\;\;i=1,..3;\label{3.9}
\end{equation}
where $\Pi_{0}$ and $\Pi_{i}$ are the canonical momenta conjugated to the
shift and lapso functions, respectively; $H_{c}$ is the canonical Hamiltonian
and $p_{0}$ is the ``momentum''\ conjugated to the time parameter. So, in this
approach, both $N$ and $N^{i}$ are taken as evolution parameters, together
with $t$, from the beginning. Now we apply the integrability conditions given
by Eq. (\ref{i14}) to the equations above. After some calculation we obtain
\begin{equation}
dH_{0}^{\prime}=\left\{  H_{0}^{\prime},H_{0c}^{\prime}\right\}  dt=-\left(
\int d^{3}xC(x)\right)  dt=0,\label{3.10}
\end{equation}
\begin{equation}
dH_{i}^{\prime}=\left\{  H_{i}^{\prime},H_{0c}^{\prime}\right\}  dt=-\left(
\int d^{3}xC_{i}(x)\right)  dt=0,\label{3.11}
\end{equation}
\begin{equation}
dH_{0c}^{\prime} = \left\{H_{0c}^{\prime},H_{0c}^{\prime}\right\}dt +
\left\{H_{c}^{\prime},H_{0}^{\prime}\right\}dN +
 \left\{H_{c}^{\prime},H_{i}^{\prime}\right\}dN^{i} = 
\left[\int d^{3}x {\partial \over \partial x^{i}}\left(F^{i}(x)\right)\right]dt
=0,\label{3.12}  
\end{equation}
where
\begin{equation}
F^{i}=N^{2}\left(  -\delta_{j}^{i}\partial_{m}-{\frac{1}{2}}T^{i}
\;_{mj}+T_{mj}\;^{i}\right)  \Pi^{\lbrack mj]},\;\;m,j=1,...,3;\label{3.12b}
\end{equation}
with the brackets on the indexes indicating antisymmetrization. Equation
(\ref{3.12}) above can be transformed in a surface integral and, therefore, at
the surface of integration we must have
\begin{equation}
F^{i}=N^{2}\left(  -\delta_{j}^{i}\partial_{m}-{\frac{1}{2}}T^{i}
\;_{mj}+T_{mj}\;^{i}\right)  \Pi^{\lbrack mj]}=0.\label{3.13}
\end{equation}

However, the latest equation  is a consequence of the fact that the canonical momenta
$\Pi^{mj}$ in this approach are symmetrical, so the antisymmetric components
$\Pi^{\lbrack mj]}$ must vanish, therefore the Equation (\ref{3.13}) is satisfied in 
in the whole spacetime, what corresponds to primary constraints in
the approach of reference \cite{M94}.

The integrability conditions given by equations (\ref{3.10}), (\ref{3.11}) and
(\ref{3.12}) imply that $C=0,$ $C_{i}=0$\ and $\Pi^{\lbrack mj]}=0$ and no new
conditions arise. The integrability conditions are equivalent to the
consistency conditions obtained in reference \cite{M94} using Dirac's method,
and the quantities $C,$\ $C_{i},$\ $\Pi^{\lbrack mj]}$ constitute a set of
first class constraints.

\section{Final remarks}

In this work we analyzed the Lagrangian density of TEGR, with a specific
choice of reference frame, by using the Hamilton-Jacobi formalism, which was
recently developed to treat singular systems. Such Lagrangian has already been
studied through Dirac's Hamiltonian formalism, where the consistency
conditions produce a set of first class constraints \cite{M94}. In our
analysis, the integrability conditions of Hamilton-Jacobi formalism produce
results that are identical to those obtained in reference \cite{M94} through
the use Dirac's Hamiltonian method.

However, one of the most interesting characteristics of Hamilton-Jacobi
formalism is the possibility of avoiding specific choices of gauge or
reference systems. So, we believe to be possible to study the Lagrangian
density of TEGR without assuming any \textit{a priori} restriction on the
tetrad fields, like Schwinger's time gauge. Our expectation is that such
restrictions should naturally arise as consequence of integrability conditions
in Hamilton-Jacobi formalism, as happens in other singular systems
\cite{G6,G7}. This question is presently under our study.

\section{Acknowledgements}

J. F. da Rocha-Neto would like to thank Professor J. Geraldo Pereira for his
hospitality at the Instituto de F\'{\i}sica Te\'{o}rica IFT/UNESP and FAPESP
(grant number 01/00890-9) for full support. P. J. Pomp\'{e}ia thanks CNPq for
full support and B. M. Pimentel thanks CNPq and FAPESP (grant number
02/00222-9) for partial support.

\end{document}